\documentclass[10pt]{article}
\usepackage{fancyhdr}
\usepackage{extramarks}
\usepackage{amsmath}
\usepackage{amsthm}
\usepackage{amsfonts}
\usepackage{siunitx}
\usepackage{tikz}
\usepackage[plain]{algorithm}
\usepackage{algpseudocode}
\usepackage{multirow}
\usepackage{booktabs}
\usepackage{authblk}
\usepackage{graphicx}
\usepackage{subfigure}
\usepackage[colorlinks,linkcolor=black,anchorcolor=black,citecolor=black,urlcolor=blue]{hyperref}
\usepackage{longtable}
\usepackage{color}
\usepackage{amsmath,bm}
\usepackage{booktabs}
\usepackage{mathtools}
\usepackage{amssymb}
\usepackage{caption}
\usepackage{capt-of}
\usepackage{mciteplus}
\usepackage{cite}
\usepackage{lipsum} 
\usepackage{mathrsfs}
\usepackage[title,titletoc,toc]{appendix}
\usepackage{xr}
\usepackage{parskip}
\usepackage{soul}
\usepackage{textcomp}
\usepackage[colaction]{multicol}
\usepackage[switch]{lineno}
\usepackage{lipsum}
\usepackage{etoolbox}
\usepackage{longtable}
\usepackage{array}
\usepackage{tablefootnote}
\usepackage{ragged2e}
\newcolumntype{C}[1]{>{\centering\arraybackslash}p{#1}}
\usetikzlibrary{automata,positioning}
\usepackage{adjustbox}
\usepackage{makecell}
\usepackage{pdflscape}
\usepackage{setspace}
\usepackage{multicol}
\usepackage[utf8]{inputenc}
\usepackage[T1]{fontenc}
\usepackage{textgreek}
\usepackage{xcolor}
\usepackage{url}

\usetikzlibrary{automata,positioning}

\begin{document}
\title{Transformer-Based Characterization of Criticality in Directed Percolation} 

\author[1]{Bozheng Dou}
\author[1,*]{Wei Li}

\affil[1]{Key Laboratory of Quark and Lepton Physics (MOE) and Institute of Particle Physics, Central}

\maketitle

\abstract{Directed percolation (DP) provides a canonical framework for studying nonequilibrium phase transitions into absorbing states, yet extracting critical behavior directly from high-dimensional spatiotemporal configurations remains challenging. Here, we investigate $(1+1)$- and $(2+1)$-dimensional bond DP using a supervised Transformer encoder. Supervised learning results demonstrate that the Transformer can distinguish between the active and absorbing phases of directed percolation with high accuracy. The critical point and critical exponents estimated from the model's classification probabilities align closely with reference values from existing literature, indicating that the model not only achieves phase classification but also yields outputs containing information related to critical scaling behavior. To probe the model's internal representation, we employ attention rollout and define an attention center that quantifies the temporal centroid of the cumulative attention distribution. The attention patterns and the attention center reorganize systematically near the critical region and exhibit pronounced finite-size dependence. Finally, we examined the model's transferability between configurations generated from fully occupied initial states and those from single-seed initial states. The results indicate the presence of transferable information between the two types of data, though the transfer performance exhibits directional asymmetry. Furthermore, when using an equal amount of target-domain data, fine-tuning yielded only a marginal improvement in accuracy compared to training from scratch.} 


\maketitle

\section{Introduction}

Nonequilibrium phase transitions and critical phenomena are ubiquitous in reaction–diffusion processes, epidemic spreading, catalytic reactions, and complex-network dynamics \cite{hinrichsen2000non,odor2004universality}. Generally, compared with equilibrium phase transitions, nonequilibrium phase transitions don't satisfy detailed balance. Moreover, their steady state properties and dynamical evolution can't be described based on the principle of free-energy minimization\cite{Marro_Dickman_1999}. Therefore, identifying phase transition points from complex spatiotemporal configurations, extracting critical scaling behavior etc. remain important problems in statistical physics. Particularly, directed percolation (DP) is one of the canonical models for studying nonequilibrium phase transitions. It describes an irreversible transition between an active state and an absorbing state. Besides, DP represents a well-known universality class of absorbing-state transitions in the absence of special features such as additional symmetries, conservation laws, or long-range interactions \cite{janssen1981nonequilibrium,grassberger1981phase}. Another important point is that near the critical point, both the spatial and temporal correlation lengths diverge and obey an anisotropic dynamical scaling relation. Although Monte Carlo simulations and finite-size scaling analyses can determine critical points and critical exponents with relatively high accuracy, automatically extracting effective features associated with critical dynamics from raw configuration data remains challenging, particularly for high-dimensional systems, complex microscopic rules, or large-scale spatiotemporal configurations \cite{shen2021supervised,RevModPhys.91.045002}.

Over the past decade, machine learning has provided new and efficient approaches to the study of phase transitions. Methods such as supervised learning, principal component analysis, convolutional neural networks(CNN), and fully convolutional networks have been applied to phase classification, critical-point detection, and critical-exponent estimation in nonequilibrium systems \cite{zhang2019machine,shen2022transfer,qi2025principal}. For nonequilibrium models such as DP, other researchers' studies have shown that neural networks can learn the statistical differences between the absorbing and active phases from finite-size configurations and, when combined with finite-size scaling, recover reasonable estimates of critical parameters. For example, in standard CNNs, information from widely separated time steps is combined through successive convolutional layers, whereas self-attention directly connects temporal tokens at different positions \cite{luo2016understanding,wang2018non}. By contrast, self-attention can directly connect temporal tokens at different time steps, facilitating the modeling of long-range temporal correlations \cite{vaswani2017attention}. However, as a black box, neural-network classification results don't automatically correspond to interpretable physical mechanisms. We need validat the model’s internal representations through dedicated attribution or attention analyses \cite{ponte2017kernel,casert2019interpretable}.

The Transformer builts around the self-attention mechanism, can directly establish connections between any two positions in a sequence and is therefore particularly well suited to processing spatiotemporal data with long-range dependencies\cite{vaswani2017attention}. Originally developed for natural language processing, Transformers have since found widespread applications in fields such as drug design \cite{chen2023machine,liu2025topological}, computer vision \cite{zhu2023biformer,patro2025spectformer}, and complex networks \cite{chen2024identifying,zhang2024graph}. Near the DP critical point, this motivates the question of whether: Can a Transformer extract criticality-related features from the spatiotemporal configurations of DP? In recent years, the ability of Transformers to model global correlations has also attracted increasing attention in physics. For example, GPhyT demonstrated that Transformer-based models can infer physical dynamics from spatiotemporal simulation data and generalize to previously unseen systems \cite{wiesner2025towards}. Complementarily, some theoretical work has related Transformer architectures to statistical mechanical models, including spin-glass systems and $O(N)$ models \cite{li2025spin,sun2025phase}. These studies motivate the use of Transformers not only as flexible tools for analyzing physical data, but also as models whose internal representations may be investigated from a statistical-physics perspective. Nevertheless, high classification accuracy alone does not establish which features of a spatiotemporal configuration are used by a neural network. In studies of phase transitions, an interpretable analysis is particularly important because it can help distinguish physically relevant collective features from incidental correlations in the training data \cite{beach2018machine,casert2019interpretable}. Fortunately, the increasing use of deep neural networks in scientific prediction has also motivated the development of interpretable-learning methods. For instance, within Transformer-based models, weighted embedding mechanisms can provide an input-dependent characterization of feature importance and have been used to assist the interpretation of complex-trait phenotype prediction in crops \cite{li2026leveraging}. More generally, the SHAP framework provides a unified additive feature-attribution method for explaining individual predictions of machine-learning models \cite{lundberg2017unified}.

In this work, we investigate \((1+1)\)-dimensional and \((2+1)\)-dimensional DP models using a supervised classification framework based on a Transformer encoder. We represent the DP evolution as a temporal sequence, in which the spatial configuration at each time step is treated as a token and fed into the model. A masking mechanism is used to handle variable-length trajectories resulting from entry into the absorbing state. First, we estimate the pseudocritical points, the critical point in the thermodynamic limit, and the associated critical exponents using the model’s phase-classification probabilities in conjunction with finite-size scaling analysis. We then examine the connection between the model’s internal representations and the critical dynamics of DP. Specifically, we employ attention rollout to analyze the cumulative propagation of attention among temporal tokens and define an attention center to quantify the temporal centroid of the model’s attention. 

Our main results are as follows. First, the Transformer can distinguishes active and absorbing state DP configurations with high accuracy and the obtained critical points are close to the theoretical results. Second, the classification probabilities exhibit finite-size scaling behavior across different system sizes. Then the thermodynamic-limit critical point and the spatial correlation-length exponent were estimated. Third, this paper introduces the "attention center" and finds that it encapsulates scale information related to critical behavior. It should be emphasized that the attention center is an interpretable metric constructed from the model’s internal representations rather than a physical order parameter in the conventional sense. In total, this work provides a viable approach for using Transformers to characterize nonequilibrium critical phenomena and for relating internal deep-learning representations to dynamical scaling behavior.

\section{Methods}\label{Method}

\subsection{DP}

In this study, we mainly apply Transformer techniques to \((1+1)\)-dimensional and \((2+1)\)-dimensional directed bond percolation (bond DP) models. The system is defined on a spatiotemporal lattice consisting of \(d\) spatial dimensions and one temporal dimension \(t\). Its dynamics follow the rules below:

\begin{figure*}[!b]
    \centering
    \subfigure{
    \includegraphics[width=15cm]{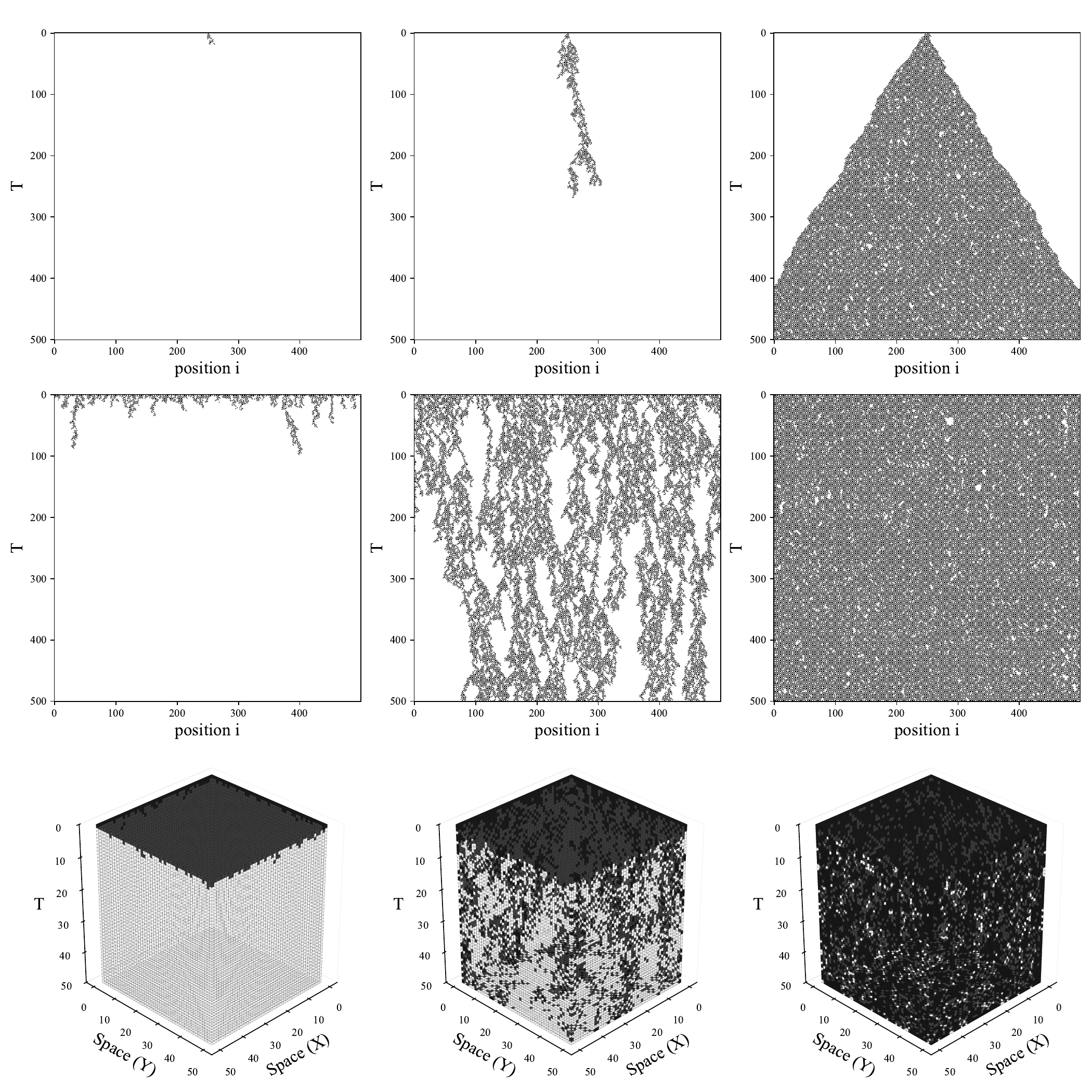}}
    \begin{minipage}{0.89\textwidth}
        \caption{\((1+1)\)- and \((2+1)\)-dimensional DP. The first and second rows show the evolution of \((1+1)\)-dimensional DP initialized from a single active seed and a fully occupied top row, respectively. The system size is \(L=500\), and the bond probabilities are \(p=0.55\), \(0.6447\), and \(0.80\). The third row shows the evolution of \((2+1)\)-dimensional DP initialized from a fully occupied top layer. The system size is \(L=50\), and the bond probabilities are \(p=0.10\), \(0.2873\), and \(0.40\), respectively.}
        \label{fig1}
    \end{minipage}
\end{figure*}

State space: The state \(s_i(t)\) of each lattice site \(i\) at time \(t\) takes the value 0 (vacant) or 1 (occupied). If all lattice sites in a given time slice are in state 0, the system is in the absorbing state. Once the system enters this state, it cannot leave spontaneously.

Evolution rule: Whether site \(i\) is occupied at time \(t+1\) depends on the states of its neighboring sites at time \(t\) and on the occupation probability \(p\).

For $(1+1)$-dimensional bond directed percolation, the state of site $i$ at time $t+1$ is determined by its two parent sites at time $t$:

\begin{equation}
s_i(t+1)=
\begin{cases}
1, &
\bigl[s_{i-1}(t)=1 \ \land \
z_{i-1\rightarrow i}(t)<p\bigr]
\ \lor \
\bigl[s_{i+1}(t)=1 \ \land \
z_{i+1\rightarrow i}(t)<p\bigr],\\[4pt]
0, & \text{otherwise}.
\end{cases}
\label{eq:bond_dp_update}
\end{equation}

\noindent where $s_i(t)\in\{0,1\}$ denotes the state of site $i$ at time $t$. Of course, periodic boundary conditions are present in every DP configuration. It should be noted that two types of initial conditions are considered for bond DP: a fully occupied initial state and a single-site occupied initial state. Both initial states evolve according to the rules described above, and representative evolution patterns under different conditions are shown in Figure.\ref{fig1}.

Bond DP exhibits critical behavior and obeys scaling hypotheses. At the critical point, the density of active sites decays algebraically from a fully occupied initial state,

\begin{equation}
    \rho(t)\propto t^{-\alpha}.
\end{equation}

\noindent where $\alpha$ is the critical decay exponent of the activity density. Below the critical point, the activity density eventually decays exponentially toward the absorbing state as shown in the first column of Figure. 1. Above the critical point, a nonzero stationary density is sustained in the thermodynamic limit as illustrated by the third column of Figure.1. Near the critical point, the spatial and temporal correlation lengths diverge with generally different exponents:

\begin{equation}
    \xi_{\perp} \propto \left|p-p_c\right|^{-\nu_{\perp}},
    \qquad
    \xi_{\parallel} \propto \left|p-p_c\right|^{-\nu_{\parallel}}.
\end{equation}

The dynamic exponent is defined as $z=\nu_{\parallel}/\nu_{\perp}$. It characterizes the spatiotemporal coupling of perturbation propagation. Subsequently, to accurately determine the critical properties of the system in the thermodynamic limit, we employ finite-size scaling theory\cite{hinrichsen2006non}. We performed Monte Carlo simulations for systems of different linear sizes \(L\) (e.g., \(L=8,16,\ldots,48\)). In a finite system, Different values of \(L\) correspond to different relaxation times. According to the scaling hypothesis, at the critical point \(p=p_c\), the characteristic time of DP scales with the system size as \cite{hinrichsen2000non}:

\begin{equation}
\tau \sim L^z.
\end{equation}

The phase classification probability is assumed to obey the finite-size scaling form:

\begin{equation}
P_{\mathrm{act}}(p,L)
=
\mathcal{F}\!\left[(p-p_c)L^{1/\nu_\perp}\right].
\label{finize}
\end{equation}

The pseudocritical point $p_c(L)$ is defined by $P_{\mathrm{act}}[p_c(L),L]=1/2$. Then, we estimate the thermodynamic-limit critical point using the linear extrapolation
\begin{equation}
p_c(L)=p_c+\frac{a}{L}.
\label{eq:pc_extrapolation}
\end{equation}
\noindent where $a$ is a fitting coefficient and the intercept at $1/L=0$ gives the extrapolated critical point $p_c$.

\subsection{Transformer}
In order to effectively extract spatiotemporal features from generated DP configuration, our research uses a deep learning architecture based on Transformer. Transformer abandons the traditional recurrent neural network structure and is entirely based on the attention mechanism to capture long-range correlations near critical points, which is consistent with the correlation length divergence characteristics in percolation theory. For the input feature vector sequence, three matrices are generated through linear transformation: query matrix (Q), key matrix (K) and value matrix (V). The attention weight is obtained by calculating the dot product of Q and K and scaling it:

\begin{equation}
    \text{Attention}(Q, K, V) = \text{softmax}\left(\frac{Q K^T}{\sqrt{d_k}}\right) V.
\end{equation}

\noindent where $d_k$\(d_k\) is the dimensionality of the key vectors. This mechanism allows the model to dynamically assign weights to all other positions in the sequence when processing information at the current moment or position, thereby achieving parallel aggregation of global information.

To enable the model to learn distinct representations from multiple subspaces simultaneously, we employ a multi-head attention mechanism. Multi-head attention projects \(Q\), \(K\), and \(V\) into different linear subspaces, performs attention calculations in parallel, concatenates the resulting outputs, and then applies another linear transformation:

\begin{equation}
\operatorname{MultiHead}(Q,K,V)
=
\operatorname{Concat}
\left(
\operatorname{head}_1,\ldots,\operatorname{head}_n
\right)W^O.
\end{equation}

This parallel architecture substantially enhances the model’s ability to capture complex spatiotemporal correlations. However, Transformer contains neither recurrent nor convolutional structures, so it is inherently insensitive to the positional order of the input sequence. To incorporate the temporal ordering information of the configurations, positional encoding is added at the input embedding layer. A combination of sine and cosine functions is commonly used to encode absolute or relative positional information:

\begin{align}
\mathrm{PE}_{(\mathrm{pos},\,2i)}
&=
\sin\left(
\frac{\mathrm{pos}}
{10000^{2i/d_{\mathrm{model}}}}
\right),\\
\mathrm{PE}_{(\mathrm{pos},\,2i+1)}
&=
\cos\left(
\frac{\mathrm{pos}}
{10000^{2i/d_{\mathrm{model}}}}
\right).
\end{align}

\begin{figure*}[!t]
    \centering
    \subfigure{
    \includegraphics[width=15cm]{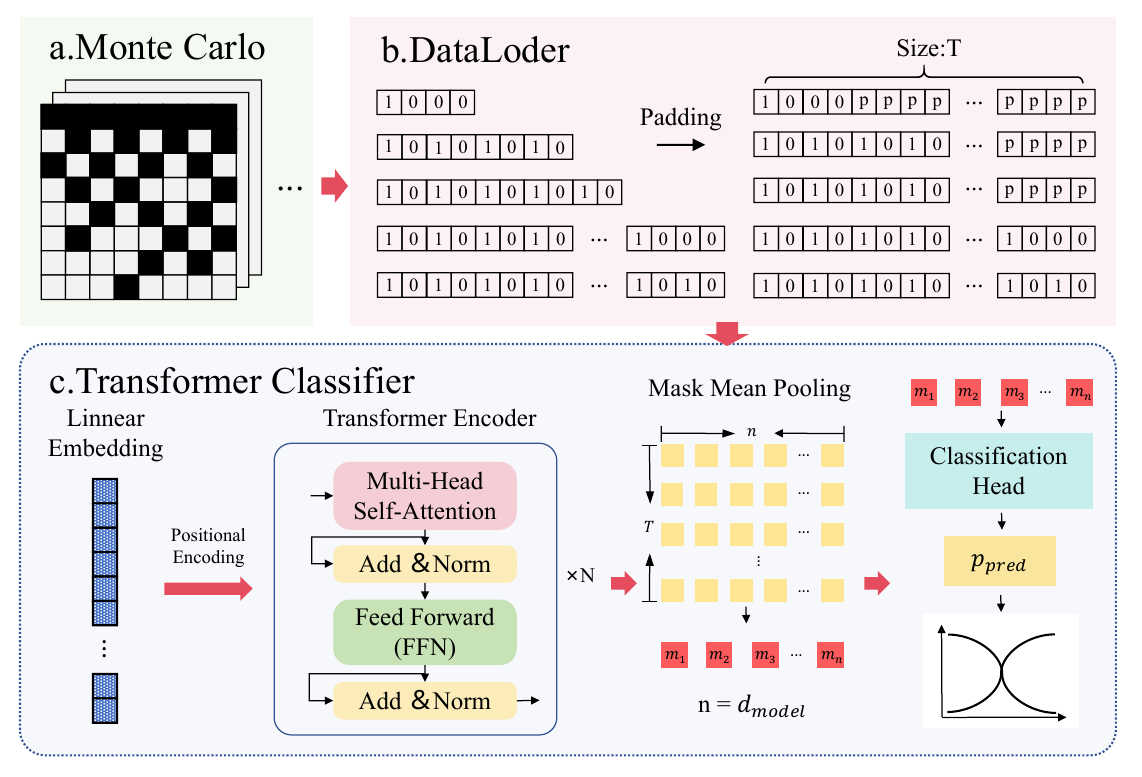}}
    \begin{minipage}{0.89\textwidth}
        \caption{Workflow of Transformer-based supervised learning.a, Monte Carlo generation of directed-percolation configurations.b, Preprocessing of the directed-percolation configuration data. For a configuration of a given system size, if its evolution does not reach the corresponding maximum number of time steps, it is padded to the required sequence length. Configurations that already reach the maximum temporal extent \(T\) are left unchanged. c, Transformer-based classification. The linear embedding module processes all samples in batches. After positional encodings are assigned to the tokens, the tokens are passed through the Transformer encoder and mapped to high-dimensional representations. A masked mean-pooling layer then aggregates these representations into a mean-pooled vector. This vector is passed through the classification head to predict the probability that the configuration percolates. The mean predicted probability can likewise be obtained for each bond probability.}
        \label{fig2}
    \end{minipage}
\end{figure*}

\noindent Where \(\mathrm{pos}\) denotes the position of a token in the sequence, \(i\) denotes the dimension index, and \(d_{\mathrm{model}}\) is the dimensionality of the model’s embedding space. Because of the unique architecture of the Transformer, each temporal token represents the full spatial configuration at one time step. For \((2+1)\)-dimensional DP, the spatial configuration is flattened before linear embedding. The model architecture consists of \(N\) encoder layers, each containing a multi-head self-attention sublayer and a fully connected feed-forward neural network. Further, Residual connections and layer normalization are incorporated to ensure stable training of the deep network. The Transformer architecture described above is illustrated in Figure. \ref{fig2}.

We then employ supervised learning to classify finite-size directed-percolation configurations, with the aim of accurately estimating the critical point \(p_c\) and the associated critical exponents. To this end, we construct a Transformer encoder model capable of processing two-dimensional configuration data with spatiotemporal correlations. The encoder consists primarily of the following components.

Notably, because the system may cease evolving once it enters the absorbing state, different samples can have different temporal lengths. To efficiently process variable-length sequences during batch training, we introduce a padding mask that allows the Transformer’s self-attention mechanism to ignore invalid padded time steps. Masked mean pooling is then used to extract the global features. Specifically, the model averages only over valid, non-padded time steps, as expressed by the following equation:

\begin{equation}
h_{\mathrm{sys}}
=
\frac{\sum_{t=1}^{T} x_t M_t}
{\sum_{t=1}^{T} M_t+\varepsilon}.
\end{equation}

\noindent where \(M_t\) is the masking matrix and \(\varepsilon\) is a small positive constant introduced to prevent division by zero. The aggregated global feature vector \(h_{\mathrm{sys}}\) is then passed through a fully connected layer, which outputs a scalar logit. The model output represents the probability that a sample belongs to the active phase. Additionally, the padding mask was applied both to the self-attention operation and to masked mean pooling. Hence, time steps following absorption contributed neither to the attention scores nor to the pooled representation.

During training, the model is optimized using the binary cross-entropy loss. The loss function is defined as:

\begin{equation}
\mathcal{L}
=
-\frac{1}{M}
\sum_{i=1}^{M}
\left[
y_i\log\left(\sigma(\hat{y}_i)\right)
+
(1-y_i)\log\left(1-\sigma(\hat{y}_i)\right)
\right].
\end{equation}

\noindent where \(\sigma\) denotes the sigmoid function. The Adam optimizer is used to train the Transformer model. After training, the phase-classification results obtained at different values of \(p\) can be used to estimate the critical point \(p_c\) of the system. The Transformer model employed in this study was implemented using PyTorch 2.6.0.

\subsection{Attention Rollout and Attention Center}

After completing the phase classification of the DP configurations, we further extracted the model’s internal self-attention weights to investigate which stages of the evolution the Transformer primarily uses to distinguish the active phase from the absorbing phase. Because high-level representations combine information from multiple lower-layer tokens, examining the raw attention matrix of a single layer alone cannot fully capture the cumulative influence of the input tokens on the final representation. We therefore employed the attention rollout method proposed by Abnar and Zuidema\cite{abnar2020quantifying}. This method recursively combines the attention matrices across Transformer layers and incorporates residual connections into the attention-propagation process, thereby providing an approximate characterization of the layer-by-layer propagation paths through which information flows from the input tokens to higher-level representations.

At first, in our input representation, each time step corresponds to one token. For \((1+1)\)-dimensional DP, each temporal token contains the one-dimensional spatial configuration of length \(L\) at the corresponding time step. For \((2+1)\)-dimensional DP, the \(L\times L\) two-dimensional spatial configuration at each time step is flattened and encoded as a temporal token. Let \(A_{ij}^{(l)}\) denote the multi-head attention matrix averaged over all attention heads in the \(l\)-th Transformer encoder layer. Because Transformers incorporate residual connections, using only the original attention matrix would neglect the information propagated through the residual pathway of each token. Therefore, the identity matrix \(I\) is added to the attention matrix at each layer, followed by row-wise renormalization:

\begin{equation}
\widetilde{A}_{ij}^{(l)}
=
\frac{A_{ij}^{(l)} + I_{ij}}
{\sum_{j}\left(A_{ij}^{(l)} + I_{ij}\right)}.
\end{equation}

The normalized attention matrices are then successively multiplied from the shallowest layer to the deepest layer to obtain the final attention rollout matrix:

\begin{equation}
\mathbf{R}
=
\widetilde{\mathbf{A}}^{(N)}
\widetilde{\mathbf{A}}^{(N-1)}
\cdots
\widetilde{\mathbf{A}}^{(1)}.
\end{equation}

\noindent where \(N\) denotes the total number of Transformer encoder layers. The matrix element \(R_{ij}\) represents the cumulative attention relationship between the \(i\)-th output temporal token and the \(j\)-th input temporal token after attention has propagated through multiple layers. Compared with examining the attention matrix of a single layer, attention rollout simultaneously incorporates indirect propagation paths across multiple layers and is therefore more suitable for characterizing the model’s attention distribution throughout the entire temporal evolution. In this work, rollout heatmaps are used to compare the temporal attention patterns of DP configurations in the absorbing phase, the critical region, and the active phase.

To further quantify how attention rollout varies with the control parameter \(p\), we define the attention center based on the rollout matrix. Because the classifier ultimately uses the global mean representation of all valid temporal tokens, the rollout matrix is first summed along the query dimension, reducing the two-dimensional attention matrix to the cumulative attention weight assigned to each input time position:

\begin{equation}
q_j = \sum_i R_{ij}.
\end{equation}

The resulting weights are then normalized as follows:

\begin{equation}
w_j
=
\frac{q_j}{\sum_j q_j},
\qquad
\sum_j w_j = 1.
\end{equation}

Here, \(w_j\) represents the relative weight assigned to the \(j\)-th temporal position throughout the multilayer attention-propagation process. Using the temporal-token index as the positional coordinate, the attention center is defined as the first moment of the attention distribution:

\begin{equation}
C_{\mathrm{att}}
=
\sum_{j=0}^{T_{\mathrm{eff}}-1} j w_j.
\end{equation}

\noindent Where \(T_{\mathrm{eff}}\) denotes the effective evolution length of a sample. A smaller \(C_{\mathrm{att}}\) indicates that the model’s cumulative attention is concentrated primarily at earlier stages of the evolution, whereas a larger \(C_{\mathrm{att}}\) indicates that the center of attention shifts toward later temporal regions. For configurations that have already entered the absorbing state, all subsequent all-zero time steps are excluded from both the model’s attention calculation and the statistical evaluation of the attention center through an attention mask. This prevents invalid temporal positions after extinction from affecting the attention centroid.
For each fixed value of \(p\), \(C_{\mathrm{att}}\) is further averaged over all samples in the test set generated at that value of \(p\):

\begin{equation}
\overline{C}_{\mathrm{att}}(p)
=
\frac{1}{N_p}
\sum_{n=1}^{N_p}
C_{\mathrm{att}}^{(n)}.
\end{equation}

This yields the dependence of the attention center on the control parameter \(p\). By comparing the evolution of \(\overline{C}_{\mathrm{att}}(p)\) across different system sizes, we can further investigate whether the Transformer’s internal attention distribution undergoes a systematic reorganization near the critical region of DP and whether it exhibits finite-size dependence.

\section{results}\label{result}

\subsection{Data and Embedding}
First, directed-percolation configurations of different sizes were generated using Monte Carlo simulations. The system sizes considered were \(L=4, 8, 16, 32\) and \(48\). The temporal extent \(t\) was determined from the scaling relation between \(L\) and \(t\), using \(z=1.58\) for \((1+1)\)-dimensional DP and \(z=1.76\) for \((2+1)\)-dimensional DP\cite{PhysRevE.108.044801}. Next, for $(1+1)$-dimensional DP, 41 bond probabilities were sampled uniformly over the interval $[0.4,0.9]$. For $(2+1)$-dimensional DP, the bond probabilities were sampled over $[0,0.5]$. Moreover, for each bond probability, 2,500 samples were generated and labeled by comparing the corresponding bond probability with the theoretical critical bond probability. Importantly, configurations generated below the reference critical point, $p<p_c^{\rm ref}$, were assigned the absorbing-phase label $y=0$, whereas those generated above it, $p>p_c^{\rm ref}$, were assigned the active-phase label $y=1$. The training, validation, and test sets were divided in a ratio of \(5:2:2\), and all three datasets were generated using the same procedure.

The distinctive evolution rules of DP determine how its configurations are encoded for input into the Transformer. Each generated configuration was ultimately represented as a two-dimensional binary matrix containing only zeros and ones, as illustrated in Figure.\ref{fig2}(a) and (b). Notably, under the same training protocol, configurations initialized from a fully occupied time slice yielded higher and more stable test accuracies than single-seed configurations. We therefore used the fully occupied initial condition for the main finite-size-scaling and attention analyses, while retaining single-seed configurations for the transfer-learning experiment in Section~4.2. Furthermore, through accuracy-based evaluation, we identified the optimal Transformer hyperparameters for each system size and dimensionality, as detailed in Supplementary Material S1. The best classification accuracies are presented in Section 3.2.

\subsection{Phase Classification and Finite-size Scaling}

We trained the model on the training set and compared its predictions with the existing labels on the validation set. Once the hyperparameters had been determined, the model’s accuracy was evaluated on the test set. The results of each training run are presented in Table \ref{accuracy}. Notably, configurations evolved from a fully occupied top row yielded higher classification accuracy than those evolved from a single seed. Moreover, The test accuracies exceeded 0.8 for all system sizes considered, indicating that the trained classifiers distinguish the two label-defined regimes with consistently high out-of-sample accuracy.

\begin{table}[!t]
    \centering
    \caption{Test accuracies of the Transformer classifier for bond DP configurations generated under different initial conditions, spatial dimensions, and system sizes.} \label{accuracy}
    \begin{tabular}{ccccccc}
        \toprule
        Initial Condition & $d$ & $L=4$ & $L=8$ & $L=16$ & $L=32$ & $L=48$ \\
        \midrule
        Single seed    & $(1+1)$ & 0.8193 & 0.8834 & 0.9050 & 0.9156 & 0.9177 \\
        Fully Occupied & $(1+1)$ & 0.8259 & 0.9216 & 0.9601 & 0.9802 & 0.9886 \\
        Fully Occupied & $(2+1)$ & 0.9525 & 0.9855 & 0.9945 & 0.9977 & 0.9994 \\
        \bottomrule
    \end{tabular}
\end{table}

\begin{figure*}[!t]
    \centering
    \subfigure{
    \includegraphics[width=15cm]{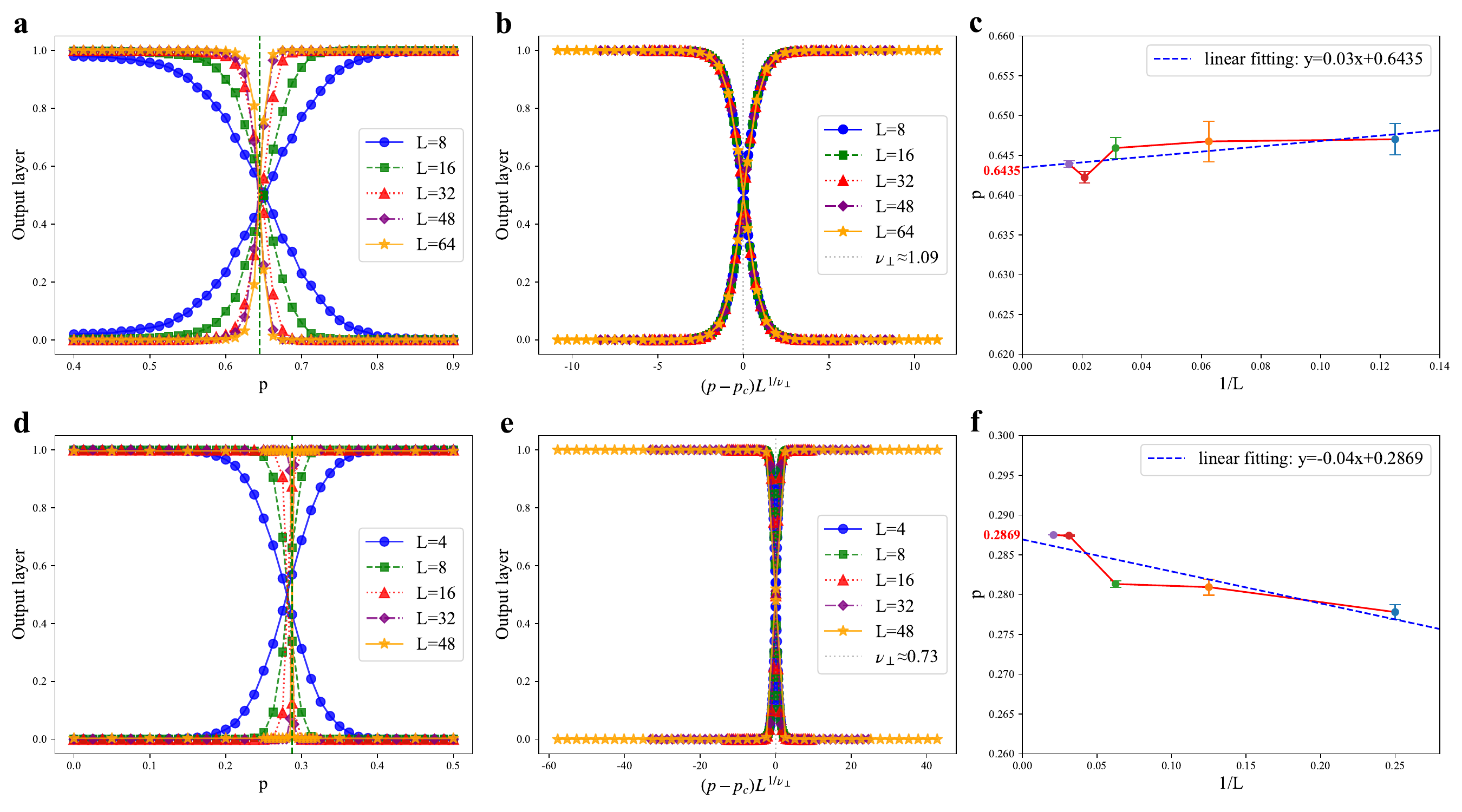}}
    \begin{minipage}{0.89\textwidth}
        \caption{Transformer results for \((1+1)\)-dimensional (top row) and \((2+1)\)-dimensional (bottom row) DP. The first column shows the critical bond probabilities estimated from finite-size configurations. The second column presents the data-collapse results, where the horizontal axis denotes the finite-size scaling variable and the vertical axis denotes the mean model output. The third column shows the critical points \(p_c\) in the infinite-size limit for both dimensionalities, estimated using finite-size scaling.}
        \label{fig3}
    \end{minipage}
\end{figure*}

During evaluation on the test set, we also predicted the percolation probability of each configuration and averaged the predicted probabilities over all configurations generated at each bond probability. The results are shown in Figure.\ref{fig3}(a) and (d). For each bond probability $p$, we averaged the predicted active-phase probability $P_{\rm act}$ over all test configurations generated at that value of $p$. The absorbing-phase probability is then given by $P_{\rm abs}=1-P_{\rm act}$. The finite-size pseudocritical point $p_c(L)$ was defined by the condition:

\begin{equation}
P_{\rm act}[p_c(L),L]=0.5.
\end{equation} 

The value of $p_c(L)$ was obtained by interpolating the two sampled bond probabilities adjacent to $P_{\rm act}=0.5$. The critical point in the thermodynamic limit can subsequently be estimated through finite-size scaling of \(p_c(L)\)\cite{carrasquilla2017machine}. Owing to finite-size effects, however, the apparent critical point shifts with system size, and no genuine phase transition occurs in a finite system. We therefore employed finite-size scaling to estimate the critical point in the infinite-size limit. According to the finite-size scaling ansatz introduced in
Section~2.1, the classification probability should obey Eq.\ref{finize} by rescaling the horizontal axis as $(p-p_c)L^{1/\nu_\perp}$, the probability curves for different system sizes collapse approximately onto a common scaling function. By combining them and adjusting the value of \(\nu_{\perp}\), data for different system sizes can be collapsed onto a common curve. The resulting combination, \((p-p_c)L^{1/\nu_{\perp}}\), is a dimensionless scaling variable and therefore defines a scale-invariant function. The results are shown in Figure.\ref{fig3}(b) and (e). As can be seen, after data collapse, the five sets of curves approximately overlap. Based on this property, we obtained \(\nu_{\perp}=1.09\pm0.02\) and \(0.73\pm0.02\) for the \((1+1)\)- and \((2+1)\)-dimensional systems, respectively.

Finally, finite-size scaling was used to determine the critical point in the infinite-size limit, as shown in Figure.\ref{fig3}(c) and (f). We plotted \(1/L\) on the horizontal axis and the estimated critical bond probability on the vertical axis. The figure shows the critical bond probabilities predicted for configurations of each finite system size. These data points can be fitted by a straight line. Consequently, when the horizontal coordinate approaches zero—that is, in the limit \(L\rightarrow\infty\)—the bond probability obtained by extrapolating the fitted line corresponds to the critical point of DP in the thermodynamic limit. As shown in the figure, the predicted critical points for \((1+1)\)- and \((2+1)\)-dimensional DP are 0.6435 and 0.2869, respectively. These values agree with the corresponding theoretical critical values of 0.6447 and 0.2873 \cite{jensen1999low,grassberger1989directed}.

\subsection{Attention rollout Analysis and Attention Center Behavior}

\begin{figure*}[!b]
    \centering
    \subfigure{
    \includegraphics[width=15cm]{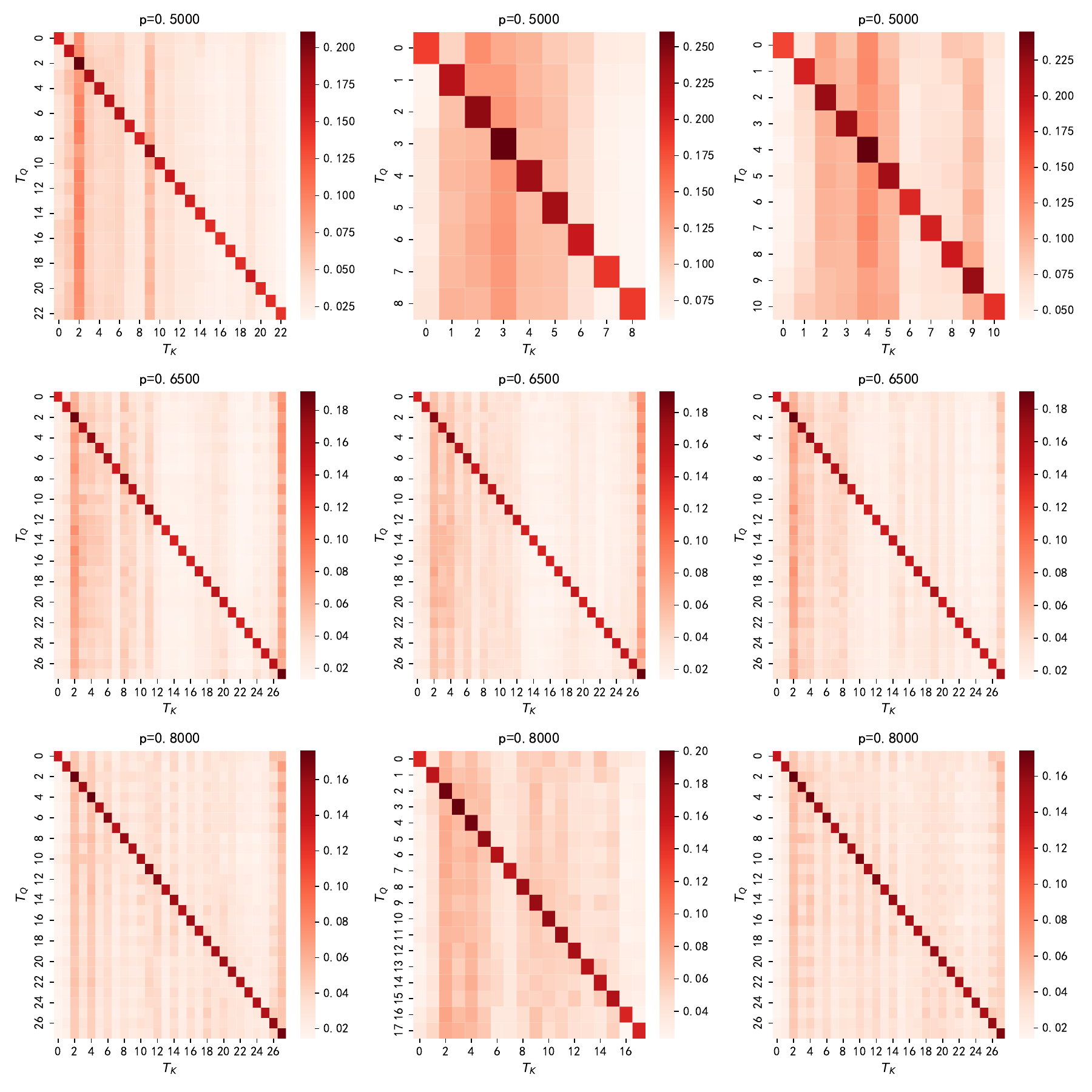}}
    \begin{minipage}{0.89\textwidth}
        \caption{Attention rollout for representative DP configurations at different bond probabilities for \(L=8\). The first, second, and third rows correspond to \(p=0.50\), \(0.65\), and \(0.80\), respectively, with three independent samples shown in each row. The horizontal axis represents the temporal positions of the key tokens, while the vertical axis represents those of the query tokens. The color intensity indicates the cumulative attention weight between two temporal tokens after attention propagation across multiple layers, with darker colors representing larger weights.}
        \label{fig4}
    \end{minipage}
\end{figure*}

To probe the temporal information used by the trained classifier, we applied attention rollout to the Transformer models evaluated on the independent test set. Because each temporal token represents the full spatial configuration at one time step, the resulting rollout matrix characterizes information propagation between different stages of the evolution rather than attention to individual spatial sites. Figure.\ref{fig4} presents the attention rollout results for representative configurations, using the \((1+1)\)-dimensional system with \(L=8\) as an example. As shown in Figure.\ref{fig4}, the pronounced diagonal contribution is partly expected from the residual connections included in attention rollout. The physically more informative features are the off-diagonal attention structures and their systematic redistribution across different dynamical regimes. In addition to the main diagonal, some samples exhibit distinct vertical regions of high attention weights. In such regions, a particular key time position receives attention from multiple query time positions, suggesting that the configuration at that time plays an important role in the model’s classification. Further comparison among the samples reveals that, for configurations that eventually enter the absorbing state, strong nonlocal attention generally occurs at relatively early temporal positions. By contrast, for configurations that remain active, larger attention weights are more frequently distributed toward the later stages of the evolution and, in some samples, are particularly concentrated within the final few valid time steps. These observations indicate that the Transformer does not rely solely on the configuration at a single time step when distinguishing different DP evolution patterns. Instead, it adjusts its attention allocation according to the temporal evolution of the system and assigns greater weight to key temporal positions containing discriminative information. Similar patterns are also observed for other system sizes and dimensionalities, as detailed in the Supplementary Material.

\begin{figure*}[!b]
    \centering
    \subfigure{
    \includegraphics[width=15cm]{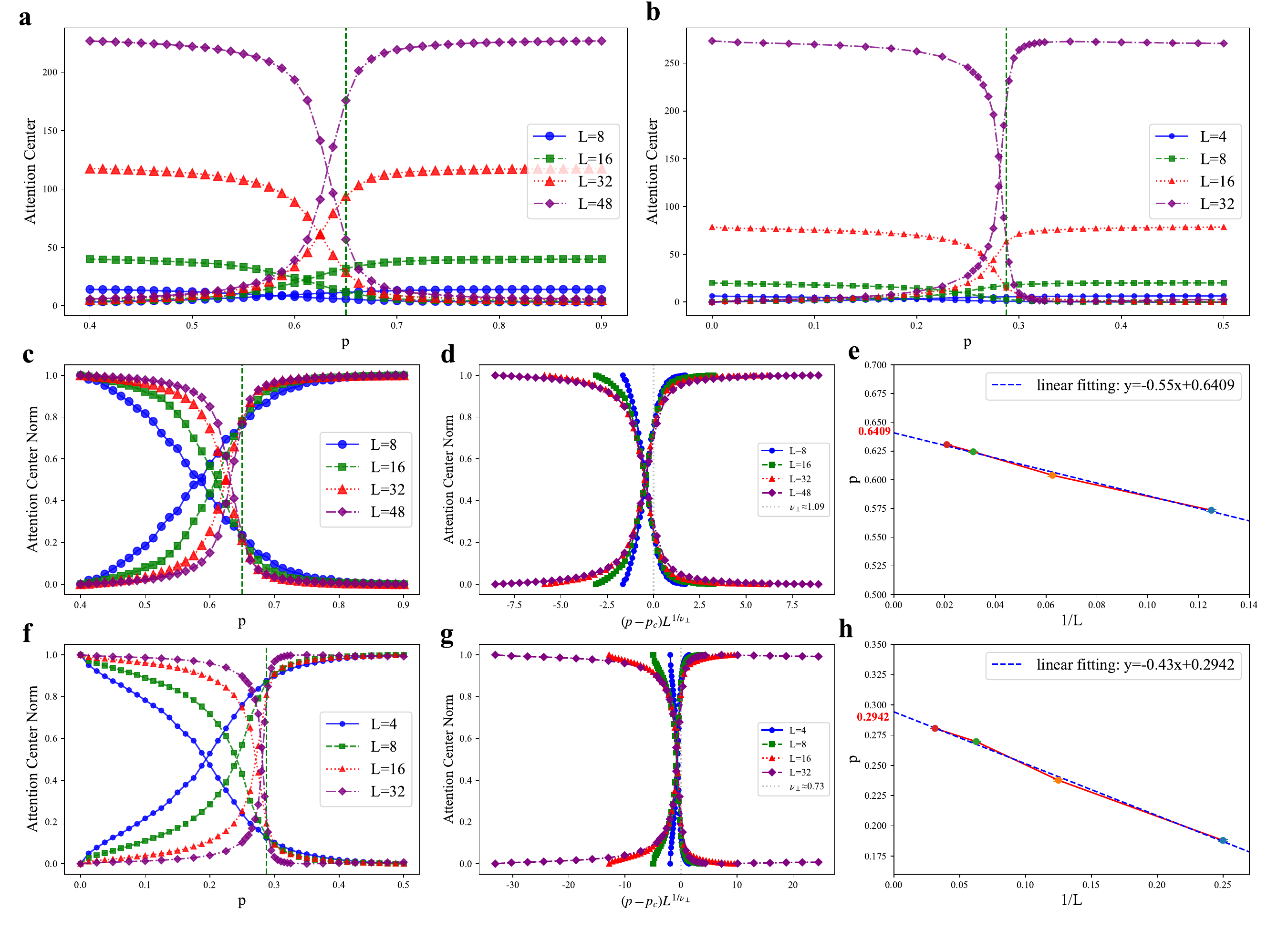}}
    \begin{minipage}{0.89\textwidth}
        \caption{Results of the Transformer attention center for DP. Panels (a), (c), (e), and (g) show the results for \((1+1)\)-dimensional DP, while panels (b), (d), (f), and (h) show those for \((2+1)\)-dimensional DP. Panels (a) and (b) show the variation of the attention center with the bond probability \(p\) for different system sizes; panels (c) and (d) show the corresponding normalized attention center; panels (e) and (f) present linear extrapolations of the pseudocritical points with respect to \(1/L\), used to estimate the critical point \(p_c\) in the thermodynamic limit; and panels (g) and (h) show the finite-size data collapse of the normalized attention center, with the scaling variable \((p-p_c)L^{1/\nu_{\perp}}\) on the horizontal axis.}
        \label{fig5}
    \end{minipage}
\end{figure*}

Although individual samples generated at the same bond probability exhibit some fluctuations, their overall attention structures are highly consistent. This indicates that, while distinguishing the active phase from the absorbing phase, the Transformer also learns internal representations associated with the temporal evolution of DP. To convert this two-dimensional attention distribution into a quantitatively comparable measure, we further define the attention center to characterize the temporal centroid of the attention rollout distribution. It can analyze variations with bond probability and system size.

The dependence of the attention center on the bond probability \(p\) for different system sizes is shown in Figure.\ref{fig5}(a) and (b). The attention center exhibits an approximately sigmoidal dependence on \(p\). In the low-\(p\) region far from the critical point, it varies relatively slowly, indicating that the model focuses primarily on the early stages of the evolution. As \(p\) approaches the critical region, the attention center increases rapidly. After the system enters the active phase and \(p\) increases further, the attention center gradually approaches a plateau, indicating a pronounced shift of the attention centroid toward later temporal positions. However, the attention center also depends on the effective trajectory length. Even a uniform attention distribution gives $C_{\mathrm{att}}=(T_{\mathrm{eff}}-1)/2$. Its increase may therefore reflect both longer trajectories and a redistribution of attention within those trajectories. More importantly, as the system size increases, the variation near the critical region becomes increasingly localized and steep, exhibiting clear finite-size dependence. These results demonstrate that the attention center compresses the complex two-dimensional attention structure encoded in the rollout matrix into a quantitatively comparable measure. Similarly, its rapid variation near the critical region also corresponds to the phase-transition behavior identified above from the classification probabilities. From the perspective of the model’s internal representations, this further suggests that the Transformer’s phase classification does not rely solely on its final output layer; instead, feature representations associated with the critical dynamics of DP have already emerged within the attention mechanism.

To compare curves with different amplitude ranges, we apply min--max normalization separately for each system size: 

\begin{equation}
\widetilde{C}_{\mathrm{att}}(p,L)
=
\frac{
\overline{C}_{\mathrm{att}}(p,L)-C_{\min}(L)
}{
C_{\max}(L)-C_{\min}(L)
},
\end{equation}

\noindent where $C_{\min}(L)$ and $C_{\max}(L)$ are the minimum and maximum values over the sampled bond probabilities at size $L$. The normalized curves are shown in Figs.~5(c) and 5(f). After normalization, the curves for different system sizes exhibit similar monotonic trends and become increasingly consistent near the critical region as the system size increases. This indicates that the variation in the attention center exhibits good scaling consistency, providing a basis for further finite-size scaling analysis. Afterwards, the normalized attention center curves were used to determine the pseudocritical point for each system size. Linear extrapolation with respect to \(1/L\) was then performed to estimate the critical point in the thermodynamic limit, as shown in Figure.\ref{fig5}(e) and (h), respectively. The resulting critical points for the \((1+1)\)- and \((2+1)\)-dimensional models are 0.6409 and 0.2942, respectively. The estimates lie near the reference critical points. This agreement indicates that the Transformer’s internal attention representations provide an approximate indication of the critical region of the system’s critical region.

Furthermore, we performed data collapse of the normalized attention center using the finite-size scaling variable, with the results shown in Figure.\ref{fig5}(d) and (g), respectively. The spatial correlation-length exponent is fixed to the value estimated from the classification probabilities in Section 3.2. This collapse therefore serves as a consistency test rather than an independent estimate of the exponent. The data for different system sizes approximately collapse onto a common scaling curve, indicating that the attention center shows an approximate finite-size-dependent scaling trend and contains scale-dependent information associated with the system’s critical behavior. Compared with the data-collapse results obtained from the classification probabilities, however, the attention center exhibits some residual deviations in the critical region, and the curves for different system sizes do not overlap perfectly. These deviations may reflect finite-size corrections, finite observation times, and the model-dependent construction of the attention center. Its finite-size scaling behavior demonstrates that the internal representations learned by the model not only support phase classification but are also closely related to the system’s critical dynamics. This point will be discussed in detail in Section 4.1.

\section{Discussion}

\subsection{Interpretation and Limitations of Attention Center}
 
This study demonstrates that the Transformer can not only classify the phases of \((1+1)\)- and \((2+1)\)-dimensional DP configurations but also extract information related to critical behavior from its internal attention representations. The supervised learning results show that the phase probabilities predicted by the model exhibit a clear crossover as a function of the control parameter \(p\), while achieving reasonably reliable classification accuracy. The model also yields reasonable estimates of the critical points and critical exponents through finite-size analysis. Furthermore, the attention rollout and attention center analyses reveal that the model’s internal attention distribution changes systematically with the dynamical state of the system. The attention rollout patterns are consistent with the model integrating information across multiple temporal stages.







In this work, each time step of a configuration is treated as a token, allowing the model to establish direct connections between any pair of temporal positions. The corresponding attention rollout captures information propagation among different time steps, which may be an important reason why the Transformer can identify the critical behavior of DP. The attention center provides a quantitative characterization of this internal temporal representation. As \(p\) increases, the attention center changes rapidly near the critical region, indicating that the centroid of the model’s attention is redistributed as the system transitions from the absorbing phase to the active phase.

It should be emphasized that the attention center is not a physical order parameter in the traditional sense. Moreover, the explanatory value of attention weights remains debated \cite{jain2019attention,wiegreffe2019attention}. In particular, attention weights don't provide rigorous causal attribution, because the model’s final output is jointly influenced by other components, including feed-forward networks, residual connections, and the output layer. Based on the results we obtained, compared with the final classification probabilities, the attention center exhibits more pronounced finite-size corrections. This difference may partly reflect how the two quantities are constructed. Specifically, the classification probability is obtained through multilayer feature extraction, global aggregation, and a nonlinear output transformation, and is directly constrained by the supervised loss. Consequently, local fluctuations among individual samples are averaged and compressed to some extent in the final output. In contrast, the attention center is the first temporal moment of the attention distribution and is therefore more sensitive to the effective survival time, key stages of the evolution, and trajectory fluctuations within the critical region. Moreover, because the attention center uses the positions of the temporal tokens as weighting coordinates, its value is jointly affected by the finite temporal window and the system size.

Because the model is trained in a supervised manner using known labels, it learns the statistical differences between configurations on opposite sides of the critical point rather than discovering the phase transition entirely without prior information. The fact that the model can recover the critical region from both its output probabilities and internal attention indicates that the learned features possess a degree of physical relevance. Nevertheless, this capability still depends on the construction of the training data, the initial conditions, and the microscopic dynamical rules. Future studies could further examine the universality of the learned representations through transfer tests involving different initial conditions, different implementations of DP, or different configuration-generation protocols.

The present analysis also has several limitations. First, the range of system sizes considered is limited, and higher-order finite-size corrections are particularly pronounced for smaller systems. Second, the attention center retains only the first moment of the attention distribution, meaning that attention distributions with different shapes may have the same center. Future work could examine additional measures, including the width and entropy of the attention distribution, the fraction of nonlocal attention, and the attention weight assigned to late-time evolution. Third, occlusion experiments, input perturbations, and gradient-based attribution methods could be used to determine whether temporal positions receiving high attention are genuinely decisive for the model’s predictions. Finally, comparisons with baseline models such as convolutional and recurrent neural networks, together with transfer learning across different configuration types, would help clarify the specific advantages of Transformers in predictive accuracy, temporal-correlation modeling, and interpretability.

Overall, this study shows that the Transformer can extract internal representations associated with phase transitions and critical scaling from the spatiotemporal configurations of DP. The finite-size behavior of the attention center further indicates that the model’s attention contains information related to changes in dynamical timescales. Although this metric is jointly affected by finite-size effects, finite observation times, and the model architecture, it provides a viable analytical approach for connecting the internal representations of deep-learning models with nonequilibrium critical dynamics.

\subsection{Cross-Initial-Condition Transfer Learning in DP}

To examine whether the Transformer learned features that are specific to a particular configuration-generation protocol, we performed transfer-learning experiments between two ensembles of \((1+1)\)-dimensional bond directed-percolation configurations generated with different initial conditions. In our result, all sites on the initial time slice were active, producing approximately rectangular spacetime configurations. As for single seed, triangular causal cone is generated . These two datasets share the same microscopic bond-DP dynamics and the same absorbing-state transition, while differing substantially in their early-time geometry, density profile, and survival statistics. The transfer experiment therefore probes the sensitivity of the learned representation to the initial condition.

\begin{table}[!t]
\centering
\caption{Cross-initial-condition transfer-learning performance at different system sizes. 
Zero-shot and fine-tuning results denote target-domain test accuracies (\%) averaged over three random seeds. 
$\Delta A_{\mathrm{5\%}}$ and $\Delta A_{\mathrm{10\%}}$ are the accuracy gains of fine-tuning over training from scratch using 5\% and 10\% target-domain data, respectively.}
\label{tab:transfer_size}
\begin{tabular}{ccccccc}
\toprule
Source $\rightarrow$ Target & $L$ & Zero-shot & $\Delta A_{\mathrm{5\%}}$ & $\Delta A_{\mathrm{10\%}}$ & Fine-tune$_{\mathrm{5\%}}$ & Fine-tune$_{\mathrm{10\%}}$ \\
\midrule
Full $\rightarrow$ Single & 8 & 86.54 & +0.30 & +0.17 & 88.29 & 88.21 \\

Full $\rightarrow$ Single & 16 & 89.13 & +0.25 & +0.10 & 90.55 & 90.55 \\

Full $\rightarrow$ Single & 32 & 90.79 & +0.10 & +0.07 & 91.51 & 91.51 \\
\midrule
Single $\rightarrow$ Full & 8 & 81.56 & +0.32 & +0.11 & 91.85 & 91.86 \\

Single $\rightarrow$ Full & 16 & 60.67 & +0.17 & +0.07 & 95.89 & 95.91 \\

Single $\rightarrow$ Full & 32 & 54.51 & +0.06 & +0.04 & 98.01 & 98.06 \\
\bottomrule
\end{tabular}\label{cross}
\end{table}

For each transfer direction, the Transformer was first trained using configurations from the source domain. Its performance was then evaluated directly on an independent test set from the target domain, which constitutes the zero-shot setting. We further fine-tuned the pretrained model using either 5\% or 10\% of the labeled target-domain training data. To assess whether pretraining offered a genuine benefit, we trained an identical model from random initialization using the same target-domain data fraction as a scratch baseline. All results were averaged over three independent random seeds, and the target-domain test set was excluded from both fine-tuning and model selection.
The zero-shot results reveal a pronounced directional asymmetry. When the model was trained using fully occupied initial configurations and directly evaluated on single-seed configurations, it retained relatively high target-domain accuracies of \(89.13\pm0.32\%\) for \(L=16\) and \(90.79\pm0.30\%\) for \(L=32\). Thus, despite the change from rectangular configurations to triangular causal structures, a substantial fraction of the learned representation remained applicable. This result suggests that the model captures some spacetime characteristics shared by the two ensembles, such as the persistence or extinction of activity and the temporal organization of active clusters.

In contrast, the reverse zero-shot transfer was much weaker. The results of training the model on a single-seed configuration and testing it on a fully populated initial configuration are shown in Table \ref{cross}. The performance at \(L=16\) and \(L=32\) was only \(60.67\pm2.59\%\) and \(54.51\pm0.98\%\), respectively. The performance results indicate that the representations learned by the model from configurations evolving from a single seed are heavily constrained by the sparse initial geometric structure. In particular, there are significant differences between the early density distributions and the final evolved shapes under these two types of initial conditions. Although these two datasets originate from the same DP dynamics, these features may alter the classifier's decisions.

But when we use a small amount of data for fine-tuning, the target-domain accuracy in both transfer directions is significantly improved. For the fully occupied \(\rightarrow\) single-seed transfer, the fine-tuned accuracies were approximately \(90.55\%\) and \(90.55\%\) for \(L=16\) using 5\% and 10\% of target-domain data, respectively, and approximately \(91.51\%\) for \(L=32\). For the reverse transfer, fine-tuning recovered accuracies above \(95\%\) for \(L=16\) and above \(98\%\) for \(L=32\). These results show that the distribution shift induced by changing the initial condition can be efficiently compensated by a limited amount of target-domain supervision.

However, the improvement of fine-tuning over training from scratch was modest. Across all investigated system sizes and target-data fractions, the mean accuracy gain ranged from \(0.04\) to \(0.32\) percentage points. Fine-tuning yielded small positive mean accuracy gains over training from scratch, although the robustness of these gains requires further assessment across random seeds. The high performance after fine-tuning is primarily attributable to the discriminative information contained in the target-domain data itself, while source-domain pretraining provides only a small additional advantage under the present architecture and training protocol.

Overall, these experiments indicate that the Transformer representation is neither completely tied to a single initial condition nor fully invariant under a change of configuration geometry. The relatively strong fully occupied \(\rightarrow\) single-seed zero-shot transfer demonstrates that some learned features are shared across the two DP ensembles. At the same time, the pronounced directional asymmetry and the small fine-tuning advantage show that the representation still retains substantial dependence on initial-condition-specific features. Since the present transfer task changes only the initial condition while retaining the same bond-DP update rule, it does not by itself establish universality across different nonequilibrium models. A stronger test of universality would examine transfer between different microscopic models within the DP universality class. Changes that alter the universality class would instead probe the limits of the learned representation. Such changes can alter the critical behavior and even lead to anomalous directed-percolation regimes \cite{brady2024anomalous}, providing a more stringent setting for testing transferable representations.

\section{Conclusion}

In this work, we investigated the use of a Transformer encoder for characterizing critical behavior in $(1+1)$- and $(2+1)$-dimensional bond DP. Each spatial configuration at a given time step and apply padding to configurations of varying lengths. This enables us to distinguish between the absorbing-state and active-state configurations of the DP process with high test accuracy. The averaged classification probabilities exhibit a finite-size crossover near the transition. Finite-size scaling of these probabilities yields estimates of the critical bond probability and the spatial correlation-length exponent that are consistent with established DP values.

We further examined the internal temporal representation of the Transformer using attention rollout. The results indicate differences in the model's attended locations between the absorbing state and the active state, suggesting a reorganization of the model's temporal information flow near the phase transition critical point. To quantify this behavior, we introduced the attention center as the temporal centroid of the rollout distribution. Its dependence on the bond probability and system size is correlated with the critical crossover and displays an approximate finite-size scaling trend. Nevertheless, the attention center is not a conventional physical order parameter. It depends on the trained network, the input tokenization, the masking procedure, and the finite observation window; attention rollout should therefore be regarded as a model-based representation analysis rather than a causal attribution of individual time steps.

Naturally, Attention Rollout and Attention Center are not the only methods for analyzing Transformer interpretability; other approaches, such as attention entropy and occlusion experiments, also exist and could be considered in future work. In this study, we found that the Attention Center metric is significantly influenced by finite-size effects. Consequently, larger system sizes and independent random seeds for training are required to more reliably quantify finite-size corrections. For instance, comparing the model against baselines—such as convolutional and recurrent models—using identical data and training protocols would help elucidate the specific role of the self-attention mechanism. Occlusion experiments, meanwhile, could verify whether temporal regions with high attention weights truly play a decisive role in classification outcomes. Finally, further testing could determine whether the model generalizes to systems belonging to the same DP universality class but governed by different microscopic rules; this would reveal whether the model has learned configuration features specific to a particular model or the critical dynamics shared across different systems. Cross-initial-condition transfer showed pronounced directional asymmetry, with stronger zero-shot performance from fully occupied to single-seed configurations than in the reverse direction. Fine-tuning provided only small mean accuracy gains over training from scratch with the same amount of target-domain data.

\bibliographystyle{unsrt}
\bibliography{ref}
\end{document}